\documentclass[aps,prl,preprint,tightenlines,%
              superscriptaddress,showpacs]{revtex4}
\def\mydate{\date{August 18, 2002}}
\def\myfiguresize{0.8\textwidth}
\def\SomeSpaceIfPreprint{\quad\\[2cm] \Large}
\def\bellelogo{\vbox to 16mm{
               \vss\hbox{\resizebox{!}{3cm}{
               \includegraphics{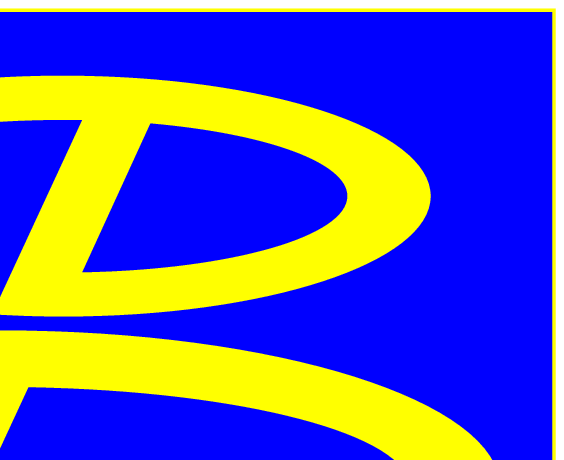}}}}\vspace{-1cm}}
\def\preprintA{\hbox{\hfil Belle Preprint 2002-28}}
\def\preprintB{\hbox{\hfil KEK Preprint 2002-83}}
\def\preprintC{\hbox{\hfil TIT-HPE-02-01}}
\def\figone{fig1_xs-ll-color.eps}
\def\figtwo{fig2_mbc-color.eps}

\usepackage{graphicx}

\def\nBB{(65.4\pm0.5)\times10^6}
\def\nBBsimple{65.4\times10^6}
\def\nLUMI{60\fbinv}

\def\signifXsee{3.4}
\def\signifXsmumu{4.7}
\def\signifXsll{5.4}

\def\CandBtoXsee{96}
\def\CandBtoXsmumu{92}
\def\CandBtoXsll{188}

\def\NumBtoXsee{25.5\,\pm11.2\,^{+4.8}_{-3.8}}  
\def\NumBtoXsmumu{37.3\,\pm9.7\,^{+7.2}_{-3.8}} 
\def\NumBtoXsll{60.1\,\pm13.9\,^{+8.6}_{-5.4}}  
\def\NumBtoXsllfull{60.1\,\pm13.9{\rm(stat)}\,^{+8.6}_{-5.4}{\rm(syst)}}

\def\BrBtoXsee{5.0\,\pm2.3\,^{+1.3}_{-1.1}}     
\def\BrBtoXsmumu{7.9\,\pm2.1\,^{+2.1}_{-1.5}}   
\def\BrBtoXsll{6.1\,\pm1.4\,^{+1.4}_{-1.1}}     
\def\BrBtoXsllFull{6.1\,\pm 1.4({\rm stat})\,^{+1.4}_{-1.1}({\rm syst})}

\def\EffBtoXsll{3.7\,\pm0.4\,\pm0.5}   
\def\EffBtoXsmumu{3.6\,\pm0.4\,\pm0.5} 
\def\EffBtoXsee{3.9\,\pm0.4\,\pm0.5}   

\def\EffElectron{92.5\%}
\def\EffMuon{91.3\%}
\def\EffKaon{90\%}
\def\FakeElectron{(0.2\pm0.06)\%}
\def\FakeMuon{(1.4\pm0.04)\%}
\def\FakeKaon{6\%}

\def\FakeXsmumu{2.6\pm0.2}
\def\FakeXsee{0.1\pm0.05}
\def\FakeXsll{2.7\pm0.2}

\def\EffMXscut{93\pm5}

\def\sysTracking{2.0\%}
\def\sysKS{8.7\%}
\def\sysPZ{6.8\%}
\def\sysElectron{1.8\%}
\def\sysMuon{2.2\%}
\def\sysKaon{2.5\%}
\def\sysPion{0.8\%}

\def\sysFisherLR{3.0\%}

\def\widthMbc{2.8}

\def\Mlla{[0.2, 1.0]}
\def\Mllb{[1.0, 2.0]}
\def\Mllc{[2.0, $M_{J/\psi}$]}
\def\Mlld{[$M_{J/\psi}$, $M_{\psi'}$]}
\def\Mlle{[$M_{\psi'}$, 5.0]}
\def\MXsa{[0.4, 0.6]}
\def\MXsb{[0.6, 0.8]}
\def\MXsc{[0.8, 1.0]}
\def\MXsd{[1.0, 1.6]}
\def\MXse{[1.6, 2.1]}

\def\BrMlla{9.3\pm7.4 \,^{+1.7}_{-1.3}}
\def\BrMllb{11.2\pm6.6\,^{+2.7}_{-2.3}}
\def\BrMllc{18.7\pm7.6\,^{+4.5}_{-3.7}}
\def\BrMlld{10.3\pm3.7\,^{+2.5}_{-2.1}}
\def\BrMlle{11.1\pm5.4\,^{+2.8}_{-2.2}}  

\def\BrMXsa{4.9\pm1.4  \,^{+1.0}_{-0.8}}
\def\BrMXsb{-0.6\pm1.0 \,^{+0.8}_{-0.5}}
\def\BrMXsc{5.2\pm3.4  \,^{+1.2}_{-1.0}}
\def\BrMXsd{26.8\pm9.5 \,^{+5.7}_{-4.6}}
\def\BrMXse{26.8\pm13.8\,^{+6.0}_{-4.4}}  

\def\epem{e^+e^-}

\def\qqbar{q\bar{q}}

\def\KS{K^0_S}

\def\pizero{\pi^0}

\def\BtoXsgamma{B\to X_s\gamma}
\def\Xs{X_s}
\def\BtoKll{B\to K\ell^+\ell^-}
\def\BtoKstarll{B\to K^*\ell^+\ell^-}
\def\BtoKorKstarll{B\to K^{(*)}\ell^+\ell^-}
\def\BtoXsll{B\to\Xs\ell^+\ell^-}
\def\BtoXsee{B\to\Xs e^+e^-}
\def\BtoXsmumu{B\to\Xs\mu^+\mu^-}
\def\BtoXsemu{B\to\Xs e^\pm\mu^\mp}

\def\GeVc{{\rm~GeV}/c}
\def\GeVcc{{\rm~GeV}/c^2}
\def\MeV{{\rm~MeV}}

\def\MeVcc{{\rm~MeV}/c^2}
\def\cm{{\rm~cm}}
\def\fbinv{{\rm~fb}^{-1}}

\def\Mbc{M_{\rm bc}}
\def\sigmabc{\sigma_{\rm bc}}
\def\DeltaE{\Delta E}
\def\Ebeam{E_{\rm beam}^{\rm CM}}
\def\pB{p_B^{\rm CM}}
\def\EB{E_B^{\rm CM}}
\def\Mll{M_{\ell^+\ell^-}}

\def\MXs{M_{X_s}}

\def\Cseven{C_7}
\def\Cnine{C_9}
\def\Cten{C_{10}}

\def\Br{{\cal B}}
\def\calL{{\cal L}}
\def\LR{{\cal LR}}
\def\Fcont{{\cal F}_{\rm cont}}
\def\Fsl{{\cal F}_{\rm sl}}

\begin{document}


\bellelogo

\preprint{\vbox{
  \preprintA
  \preprintB
  \preprintC
}}

\title{\SomeSpaceIfPreprint
Measurement of the Electroweak Penguin Process \boldmath$\BtoXsll$}

\affiliation{Aomori University, Aomori}
\affiliation{Budker Institute of Nuclear Physics, Novosibirsk}
\affiliation{Chiba University, Chiba}
\affiliation{Chuo University, Tokyo}
\affiliation{University of Cincinnati, Cincinnati OH}
\affiliation{Deutsches Elektronen--Synchrotron, Hamburg}
\affiliation{University of Frankfurt, Frankfurt}
\affiliation{University of Hawaii, Honolulu HI}
\affiliation{High Energy Accelerator Research Organization (KEK), Tsukuba}
\affiliation{Hiroshima Institute of Technology, Hiroshima}
\affiliation{Institute of High Energy Physics, Chinese Academy of Sciences, Beijing}
\affiliation{Institute of High Energy Physics, Vienna}
\affiliation{Institute for Theoretical and Experimental Physics, Moscow}
\affiliation{J. Stefan Institute, Ljubljana}
\affiliation{Kanagawa University, Yokohama}
\affiliation{Korea University, Seoul}
\affiliation{Kyoto University, Kyoto}
\affiliation{Kyungpook National University, Taegu}
\affiliation{Institut de Physique des Hautes \'Energies, Universit\'e de Lausanne, Lausanne}
\affiliation{University of Ljubljana, Ljubljana}
\affiliation{University of Maribor, Maribor}
\affiliation{University of Melbourne, Victoria}
\affiliation{Nagoya University, Nagoya}
\affiliation{Nara Women's University, Nara}
\affiliation{National Kaohsiung Normal University, Kaohsiung}
\affiliation{National Lien-Ho Institute of Technology, Miao Li}
\affiliation{National Taiwan University, Taipei}
\affiliation{H. Niewodniczanski Institute of Nuclear Physics, Krakow}
\affiliation{Nihon Dental College, Niigata}
\affiliation{Niigata University, Niigata}
\affiliation{Osaka City University, Osaka}
\affiliation{Osaka University, Osaka}
\affiliation{Panjab University, Chandigarh}
\affiliation{Peking University, Beijing}
\affiliation{Princeton University, Princeton NJ}
\affiliation{RIKEN BNL Research Center, Brookhaven NY}
\affiliation{Saga University, Saga}
\affiliation{University of Science and Technology of China, Hefei}
\affiliation{Seoul National University, Seoul}
\affiliation{Sungkyunkwan University, Suwon}
\affiliation{University of Sydney, Sydney NSW}
\affiliation{Tata Institute of Fundamental Research, Bombay}
\affiliation{Toho University, Funabashi}
\affiliation{Tohoku Gakuin University, Tagajo}
\affiliation{Tohoku University, Sendai}
\affiliation{University of Tokyo, Tokyo}
\affiliation{Tokyo Institute of Technology, Tokyo}
\affiliation{Tokyo Metropolitan University, Tokyo}
\affiliation{Tokyo University of Agriculture and Technology, Tokyo}
\affiliation{Toyama National College of Maritime Technology, Toyama}
\affiliation{University of Tsukuba, Tsukuba}
\affiliation{Utkal University, Bhubaneswer}
\affiliation{Virginia Polytechnic Institute and State University, Blacksburg VA}
\affiliation{Yokkaichi University, Yokkaichi}
\affiliation{Yonsei University, Seoul}
  \author{J.~Kaneko}\affiliation{Tokyo Institute of Technology, Tokyo} 
  \author{K.~Abe}\affiliation{High Energy Accelerator Research Organization (KEK), Tsukuba} 
  \author{K.~Abe}\affiliation{Tohoku Gakuin University, Tagajo} 
  \author{T.~Abe}\affiliation{Tohoku University, Sendai} 
  \author{I.~Adachi}\affiliation{High Energy Accelerator Research Organization (KEK), Tsukuba} 
  \author{Byoung~Sup~Ahn}\affiliation{Korea University, Seoul} 
  \author{H.~Aihara}\affiliation{University of Tokyo, Tokyo} 
  \author{M.~Akatsu}\affiliation{Nagoya University, Nagoya} 
  \author{Y.~Asano}\affiliation{University of Tsukuba, Tsukuba} 
  \author{T.~Aso}\affiliation{Toyama National College of Maritime Technology, Toyama} 
% ...
\collaboration{The Belle Collaboration}\noaffiliation

\mydate

\begin{abstract} 

We report the first measurement of the branching fraction for the
inclusive decay $\BtoXsll$, where $\ell$ is either an electron or a muon
and $\Xs$ is a hadronic recoil system that contains an $s$-quark.  We
analyzed a data sample of $\nBBsimple$ $B$ meson pairs collected with
the Belle detector at the KEKB $\epem$ asymmetric-energy collider.  We
find $\Br(\BtoXsll)=(\BrBtoXsllFull)\times10^{-6}$ for dilepton masses
greater than $0.2\GeVcc$.

\end{abstract}




\maketitle


The first observation of the flavor-changing-neutral-current
(FCNC) weak decay process $\BtoKll$, recently reported by
Belle~\cite{bib:belle-kll}, opens a new window for searches for physics
beyond the Standard Model (SM)~\cite{bib:beyond-sm}.  There are no SM
first-order weak decay processes that can produce such decays; the
dominant SM processes are second-order $b$-quark to $s$-quark processes
with a virtual $t$-quark in a loop (electroweak penguin) or a box
diagram.  Non-SM processes could produce sizable modifications to the
decay amplitude due to contributions from virtual non-SM particles (such
as charged Higgs or SUSY particles) in the loop.

The decay amplitude is often described by an effective Hamiltonian in
which non-SM contributions can modify the Wilson coefficients $\Cseven$,
$\Cnine$ and $\Cten$.  The magnitude of allowable modifications of
$\Cseven$
is constrained by the measured rate for inclusive $\BtoXsgamma$ decays
\cite{bib:belle-btosgamma,bib:cleo-btosgamma,bib:aleph-btosgamma}.  The
Belle measurement of the exclusive process $\BtoKll$ has been used to
determine the best limits on non-SM contributions to $\Cnine$ and
$\Cten$ \cite{bib:ali-2001}.  The usefulness of
exclusive measurements is limited by the large theoretical
uncertainties associated with the $s$-quark to $K$ meson hadronization
process.  Since these uncertainties are not as severe for inclusive
processes, measurements of the branching fraction and the dilepton and
hadronic mass spectra for $\BtoXsll$ will provide more stringent and
less model-dependent probes for new physics. This process has not been
measured; CLEO has reported a 90\% confidence upper limit of $\Br
(\BtoXsll)<4.2\times 10^{-5}$ \cite{bib:cleo-sll}.

In this Letter, we present the results of a measurement of the branching
fraction for the inclusive decay $\BtoXsll$, where $B$ is either $B^0$
or $B^+$, $\ell$ is either an electron or a muon, and $\Xs$ is a
hadronic recoil system that contains an $s$-quark.  Here and throughout
the paper, the inclusion of charge conjugated modes is implied.  We use
a data sample collected at the $\Upsilon(4S)$ resonance with the Belle
detector at the KEKB $\epem$ asymmetric-energy collider (3.5 GeV on 8
GeV) \cite{bib:kekb-unpublished}.  This sample contains $\nBB$ $B$ meson
pairs, corresponding to an integrated luminosity of $\nLUMI$.

The Belle detector is a large-solid-angle magnetic spectrometer that
consists of a three-layer silicon vertex detector (SVD), a 50-layer
central drift chamber (CDC), an array of aerogel threshold \v{C}erenkov
counters (ACC), time-of-flight scintillation counters (TOF), and an
electromagnetic calorimeter comprised of CsI(Tl) crystals (ECL) located
inside a superconducting solenoid coil that provides a 1.5~T magnetic
field.  An iron flux-return located outside of the coil is instrumented
with resistive plate counters to identify muons (KLM).  The
detector is described in detail elsewhere~\cite{bib:belle-nim}.

We reconstruct charged particle trajectories with the CDC and SVD.
Electron identification is based on the position and shower shape of the
cluster in the ECL, ratio of the cluster energy to the track momentum
($E/p$), specific energy-loss measurement ($dE/dx$) with the CDC and the
response from the ACC.  Muon identification is based on the
hit positions and the depth of penetration into the ECL and KLM.
We require the electron's (muon's) laboratory momentum to
be greater than $0.5\GeVc$ ($1\GeVc$).  We find an electron (muon)
selection efficiency of $\EffElectron$ ($\EffMuon$) with a
$\FakeElectron$ ($\FakeMuon$) pion to electron (muon)
mis-identification probability.
Charged kaon candidates are selected by
using information from the ACC, TOF and $dE/dx$ in the CDC.  The kaon
selection efficiency is $\EffKaon$ with a pion to kaon
mis-identification probability of $\FakeKaon$.  The remaining charged
tracks are assumed to be pions.  We select $\KS\to\pi^+\pi^-$ candidates with invariant mass within $15\MeVcc$ of the
$K^0$ mass.  We impose additional $\KS$ selection criteria based on the
distance and the direction of the $\KS$ vertex and the impact parameters
of daughter tracks.  We require the charged tracks other than those used
in the $\KS$ reconstruction to have impact parameters with respect to
the nominal interaction point of less than $0.5\cm$ in the radial
direction and $3\cm$ along the beam direction.

We reconstruct photons from ECL energy clusters that have no associated
charged tracks.  The shower shape is required to be consistent with an
electromagnetic cluster and the energy to be greater than $50\MeV$.  We
reconstruct $\pi^0\to\gamma\gamma$ candidates from photon pairs with
invariant mass within $10\MeVcc$ of the nominal $\pi^0$ mass.

The $\Xs$ system is reconstructed in 18 different combinations of either
a $K^+$ or $\KS$ combined with 0 to 4 pions, of which up to one
$\pizero$ is allowed.
We combine the $\Xs$ with two oppositely charged leptons to
form a $B$ candidate.  We identify the $\BtoXsll$ signal with the
beam-energy constrained mass, $\Mbc=\sqrt{(\Ebeam/c^2)^2 - (\pB/c)^2}$,
where $\Ebeam$ and $\pB$ are the beam energy and the $B$ candidate
momentum calculated in the center-of-mass (CM) system, respectively.  We
find the average $\Mbc$ resolution is $\sigmabc=\widthMbc\MeVcc$.  The
energy difference
$\DeltaE=\EB-\Ebeam$, where $\EB$ is the CM energy of the $B$ candidate,
is combined with other variables to provide background suppression.

\begin{figure}[ht]
 \resizebox{\myfiguresize}{!}{\includegraphics{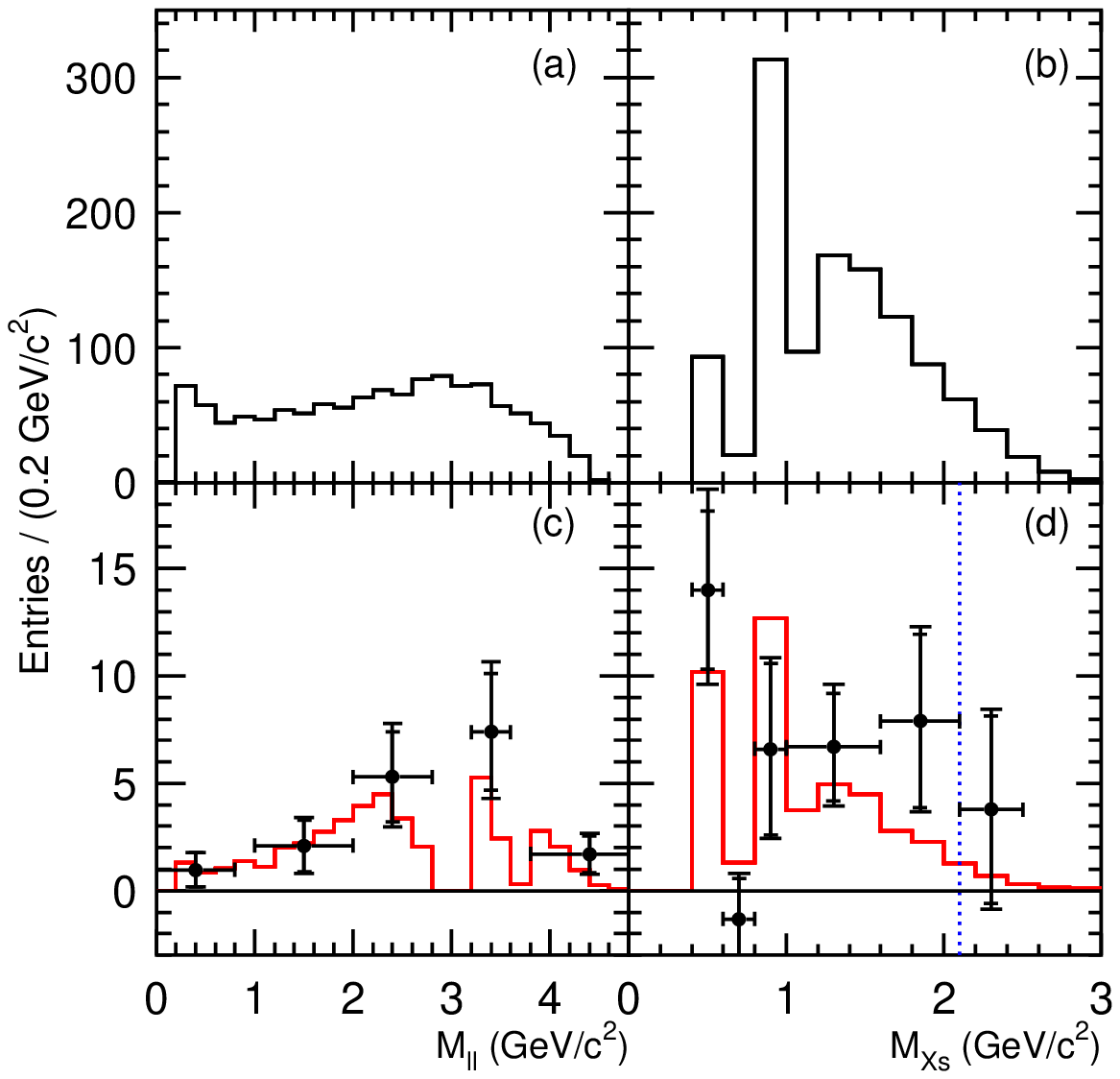}}
 \caption{SM expectations for the (a) dilepton and (b) recoil mass
 spectra; the observed (c) dilepton and (d) recoil mass
 spectra (circles).
 Inner and outer error bars indicate the statistical and total errors,
 respectively.
 The histograms in (c), (d) show the
 SM expectations after all the selections are applied;
 histograms are normalized to the expected branching fractions.  
 The gaps in (c) are due to the
 $J/\psi$, $\psi'$ vetoes.  The
 dotted line in (d) indicates the $\MXs<2.1\GeVcc$ requirement. }
 \label{fig:mll-mxs}
\end{figure}

To optimize the selection criteria and to determine their signal
efficiencies, we use the following signal model.  The SM
calculation of Ref.~\cite{bib:ali-2001} is used for the dilepton mass
($\Mll$) spectrum.  We require $\Mll$ to be greater than $0.2\GeVcc$ to
remove the virtual photon contribution, $b\to s \gamma^*\to
s\epem$~\cite{bib:ali-priv}.
We model the recoil mass ($\MXs$) spectrum using a non-resonant Fermi
motion model~\cite{bib:ali-1997} and the JETSET hadronization
program~\cite{bib:jetset}.  For $\MXs<1.1\GeVcc$, however, the spectrum
is modeled by the sum of the exclusive $\BtoKorKstarll$ components as
predicted by the SM~\cite{bib:exclusive-prediction}.  Using this model
we find that our reconstructed final states account for $81\%$ of the
total $\Xs$ states.  Figures~\ref{fig:mll-mxs}(a) and (b) show the
$\Mll$ and $\MXs$ spectra from this model.

There are two background sources that can peak in $\Mbc$ and
$\DeltaE$.  The first is hadronic $B$ decays into one kaon plus multiple
pions ($B\to \Xs\pi^+\pi^-$) from abundant decays such as $B\to
D^{(*)}n\pi$ $(n\ge1)$.  We estimate this
background contribution by reconstructing $B\to \Xs\pi^+\pi^-$ events
without the lepton identification criteria and multiplying by the
measured momentum dependent pion mis-identification probability.  We
find contributions of $\FakeXsmumu$ events to the $\BtoXsmumu$ signal
and $\FakeXsee$ events to $\BtoXsee$.  We refer to this as the fake
background, and subtract it from the signal yield.  The second is from
$B\to J/\psi \Xs$ and $B\to \psi' \Xs$ where $J/\psi$ and $\psi'$ decay
into dileptons.  These decay modes have the same final states and, in
principle, interfere with the signal.  For this study, these charmonium
decays are explicitly vetoed.  The veto windows are $-0.6$ to
$+0.2\GeVcc$ ($-0.35$ to $+0.2\GeVcc$) around the $J/\psi$ mass for the
$e^+e^-$ ($\mu^+\mu^-$) channel, and $-0.3$ to $+0.15\GeVcc$ around the
$\psi'$ mass for both channels.
We find the background from this source is negligibly small using a
$B\to J/\psi X_s$ Monte Carlo (MC) sample in which the $\Mll$ spectrum
reproduces that of data.

The largest background sources are random combinations of dileptons with
a kaon and pions that originate from continuum $\qqbar$ ($q=u,d,s,c$)
production or from semileptonic $B$ decays.  We reject 83\% of the
$\qqbar$ background with a signal efficiency of 90\% by using a Fisher
discriminant \cite{bib:fisher} ($\Fcont$) based on a modified set of
Fox-Wolfram moments \cite{bib:fox-wolfram} that differentiate the event
topology.  In the semileptonic $B$ decay background, both $B$ mesons
decay into leptons or two leptons are produced from the $b\to c\to s,d$
decay chain.  We combine the missing mass ($M_{\rm miss}$) and the total
visible energy ($E_{\rm vis}$) \cite{bib:mmiss} into another Fisher
discriminant ($\Fsl$) to reject 85\% of events with two neutrinos
with a signal efficiency of 91\%.

We further reduce the backgrounds using $\DeltaE$ and the cosine of the
$B$ flight direction ($\cos\theta_B$) with respect to the $e^-$ beam
direction in the CM frame.  First, we select events with
$|\DeltaE|<40\MeV$ $(\sim\!3\sigma_{\DeltaE})$.  We then calculate
likelihoods $\calL_{S,B} = p_{S,B}^{\DeltaE}\times p_{S,B}^{\rm cosB}$,
where $p_{S,B}^{\DeltaE}$ and $p_{S,B}^{\rm cosB}$ are the uncorrelated
probability density functions (PDF) for $\DeltaE$ and $\cos\theta_B$ for
the signal ($S$) and the background ($B$), respectively, and form a
likelihood ratio $\LR=\calL_S/(\calL_S+\calL_B)$.  We model
$p_S^{\DeltaE}$ with a Gaussian primarily determined from a signal MC
sample and calibrated using $B\to J/\psi X_s$ data; $p_B^{\DeltaE}$ is
modeled with a linear function with a slope determined from a MC
sample that contains $b\to c$ decays and $\qqbar$ events.  The signal
follows a $1-\cos^2\theta_B$ distribution, while the background
distributes uniformly in $\cos\theta_B$.  By requiring $\LR>0.8$, we
retain 75\% of the signal while removing 80\% of backgrounds.

For events with multiple candidates that pass the selection criteria, we
choose the combination with the largest value of $\LR$.  We find that
the correct candidate is reconstructed in 73\% of the events.  We reject
candidates with $\Xs$ invariant mass greater than $2.1\GeVcc$.  This
condition removes a large fraction of combinatorial background while
retaining ($\EffMXscut$)\% of the signal.

Signal MC samples are used to determine reconstruction efficiencies of
$(\EffBtoXsll)\%$, where the first error is systematic and the second
error is due to the model uncertainty.  Here, an equal production rate
is assumed for $B^0\overline{B}{}^0$ and $B^+B^-$.  Systematic errors
include uncertainties from the
tracking efficiency ($\sysTracking$ per
track), $\KS$ reconstruction efficiency ($\sysKS$ per $\KS$), $\pi^0$
reconstruction efficiency ($\sysPZ$ per $\pi^0$), electron
identification ($\sysElectron$ per electron), muon identification
($\sysMuon$ per muon), $K^+$ identification ($\sysKaon$ per $K^+$),
$\pi^+$ identification ($\sysPion$ per $\pi^+$), and the background
suppression criteria ($\sysFisherLR$ for $\Fcont$, $\Fsl$ and $\LR$,
combined, estimated using $B\to J/\psi X_s$ data).

The model uncertainty includes the following sources.  The largest error
is due to uncertainties in the SM branching fraction of the exclusive
modes (11\%).  The uncertainty in the $\MXs$ spectrum (4\%) is due to
the Fermi momentum ($p_F$) and spectator quark mass ($m_q$) parameters.
We take $p_F=0.4\GeVc$ and $m_q=0\GeVcc$, and vary the parameters over a
range allowed by the measured heavy quark effective theory (HQET)
parameters $\lambda_1$ and $\overline{\Lambda}$
\cite{bib:cleo-btosgamma,bib:cleo-hqet}.  The uncertainty in the
fraction of the unmeasured modes (2.1\%) and the uncertainty due to the
fractions of $\pi^0$ and $\KS$ contained in $\Xs$ (4.6\%) are estimated
by comparing the inclusive hadron ($\pi^0$, $\KS$, $\eta$, $\phi$, etc.)
production rates between continuum data and JETSET.

\begin{figure}[ht]
 \resizebox{\myfiguresize}{!}{\includegraphics{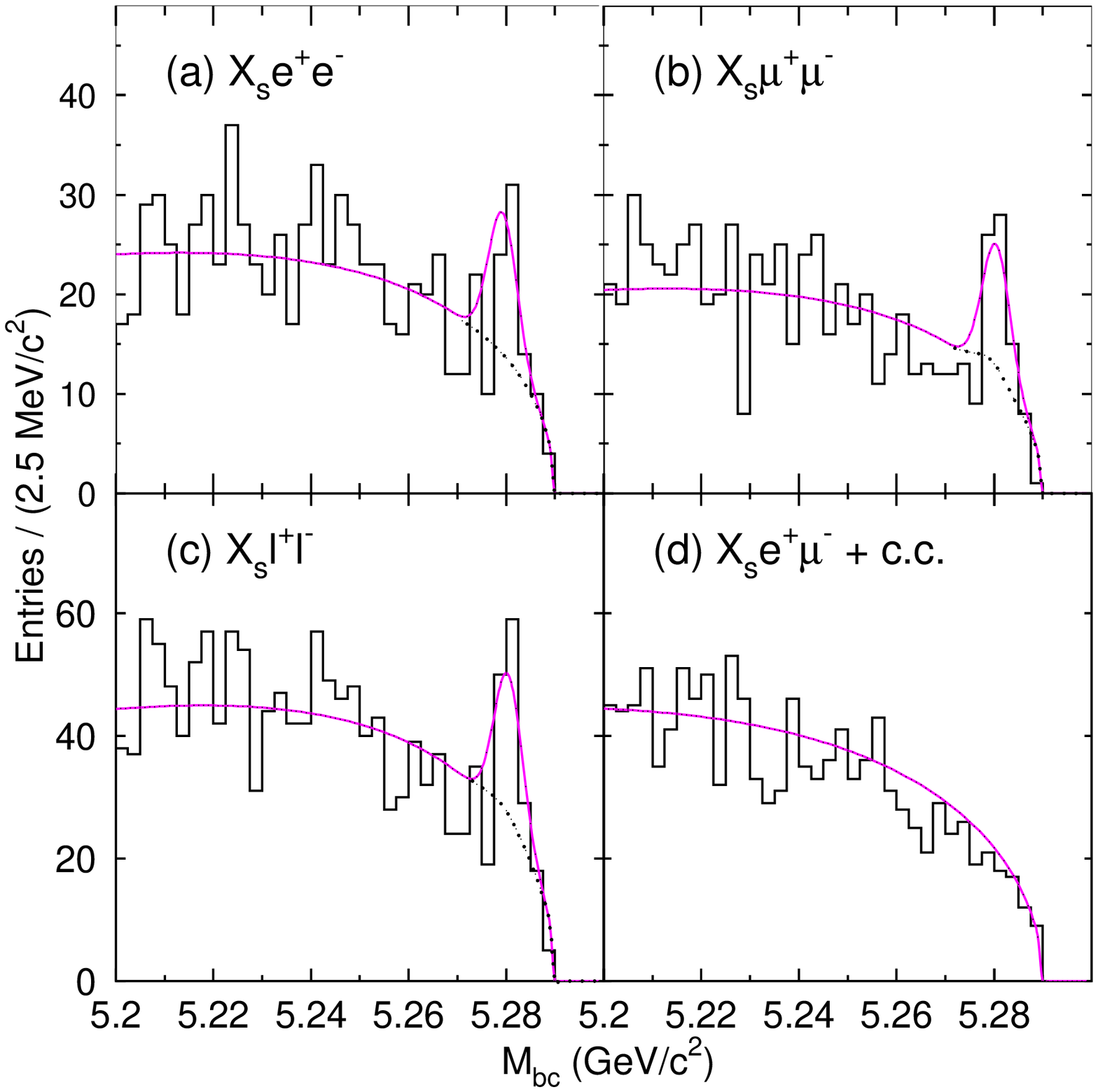}}
 \caption{Beam-energy constrained mass distributions for (a) $\BtoXsee$, (b)
 $\BtoXsmumu$, (c) $\BtoXsll$ and (d) $\BtoXsemu$.  The solid lines
 indicate the fit results and the dotted lines show the sum of the
 background components.}
 \label{fig:mbc}
\end{figure}

We determine the signal yield from an unbinned maximum likelihood fit to
the $\Mbc$ distribution as shown in Fig.~\ref{fig:mbc}.  We model the
signal with a Gaussian function and the background with a threshold
function \cite{bib:argus-function}.  The width of the signal Gaussian is
obtained from $B\to J/\psi\Xs$ data.  The background shape is obtained
from the background MC sample.  We verify that the threshold
function obtained from MC reproduces the shape of data for $B\to \Xs
e\mu$ combinations (Fig.~\ref{fig:mbc}(d)).  The fit results are
given in Table~\ref{tbl:summary}; we find $\NumBtoXsllfull$ events for
the combined $\BtoXsll$ modes (Fig.~\ref{fig:mbc}(c)) with a
statistical significance of $\signifXsll\sigma$.  Here, the systematic
error is obtained from the maximum deviation when the background
and signal shapes are varied by one standard deviation of the
statistical error in their shape parameters.  The significance is
defined as $\sqrt{-2\ln(\calL_0/\calL_{\rm max})}$, where $\calL_{\rm
max}$ is the maximum likelihood and $\calL_0$ is the maximum likelihood
when the signal yield is constrained to be zero.
The fake background and the uncertainty in the background
shape are included in the significance calculation.  We calculate the
branching fraction for $\Mll>0.2\GeVcc$ to be
\[
   \Br(\BtoXsll)=(\BrBtoXsllFull)\times10^{-6},
\]
where the systematic errors in the yield, efficiency, the number of $B$
meson pairs and the model errors are added in quadrature to give the
total systematic error.  This can be compared to the SM expectation
$\Br(\BtoXsll)=(4.2\pm0.7)\times10^{-6}$ \cite{bib:ali-priv}.
Table~\ref{tbl:summary} summarizes the branching fractions for
$\BtoXsee$ and $\BtoXsmumu$ separately, together with the number of
candidates, signal yields, fake background estimations, efficiencies and
the statistical significances.




The dilepton mass spectrum is measured by dividing the data into $\Mll$
bins.  For each bin, the signal yield is extracted from a fit to the
$\Mbc$ distribution and the fake background contribution is subtracted.
The result is shown in Fig.~\ref{fig:mll-mxs}(c).  Similarly, the recoil
mass spectrum is obtained by dividing the data into $\MXs$ bins and
extracting the signal yield for each bin.  The result is shown in
Fig.~\ref{fig:mll-mxs}(d).  With the current statistics, the $\Mll$ and
$\MXs$ spectra are in agreement with SM expectations.  Branching
fractions for each bin are given in Table~\ref{tbl:spectra}.

In summary, we present the first measurement of the inclusive branching
fraction for the electroweak penguin decay $\BtoXsll$.  The
results are in
agreement with the SM expectations and can be used to constrain
extensions of the SM.

We wish to thank the KEKB accelerator group for the excellent
operation of the KEKB accelerator.
We thank A. Ali, E. Lunghi and G. Hiller for providing us many helpful
suggestions and calculations.
We acknowledge support from the Ministry of Education,
Culture, Sports, Science, and Technology of Japan
and the Japan Society for the Promotion of Science;
the Australian Research Council
and the Australian Department of Industry, Science and Resources;
the National Science Foundation of China under contract No.~10175071;
the Department of Science and Technology of India;
the BK21 program of the Ministry of Education of Korea
and the CHEP SRC program of the Korea Science and Engineering Foundation;
the Polish State Committee for Scientific Research
under contract No.~2P03B 17017;
the Ministry of Science and Technology of the Russian Federation;
the Ministry of Education, Science and Sport of the Republic of Slovenia;
the National Science Council and the Ministry of Education of Taiwan;
and the U.S.\ Department of Energy.



\begin{table*}[ht]
\caption{Fit results for the number of candidates, signal yields,
fake backgrounds, reconstruction efficiencies, statistical
significances and branching fractions ($\Br$).  Candidates are
counted in a $\pm3\sigmabc$ window in $\Mbc$.}
\label{tbl:summary}
\begin{ruledtabular}
\begin{tabular}{cccccccc}
mode & candidates & signal yield & fake background
                  & efficiency (\%) & significance
                  & $\Br (\times10^{-6})$ \\
\hline
$\BtoXsee$   & $\CandBtoXsee$  & $\NumBtoXsee$ & $\FakeXsee$
         & $\EffBtoXsee$   &  $\signifXsee$ & $\BrBtoXsee$  \\
$\BtoXsmumu$   & $\CandBtoXsmumu$  & $\NumBtoXsmumu$ & $\FakeXsmumu$
         & $\EffBtoXsmumu$   &  $\signifXsmumu$ & $\BrBtoXsmumu$ \\
$\BtoXsll$   & $\CandBtoXsll$  & $\NumBtoXsll$ & $\FakeXsll$
         & $\EffBtoXsll$   &  $\signifXsll$ & $\BrBtoXsll$ \\
\end{tabular}
\end{ruledtabular}
\end{table*}

\begin{table}[ht]
\caption{Branching fractions ($\Br$) for each bin of $\Mll$ and $\MXs$.
  The first and second errors are statistical and systematic,
  respectively.}
\label{tbl:spectra}
\begin{ruledtabular}
\begin{tabular}{cccr}
$\Mll$ & $\Br(\times10^{-7})$ & 
$\MXs$ & $\Br(\times10^{-7})$ \\
$(\mbox{GeV}/c^2)$ & & 
$(\mbox{GeV}/c^2)$ & \\
\hline 
\Mlla & $\BrMlla$ &  \MXsa & $\BrMXsa$ \\
\Mllb & $\BrMllb$ &  \MXsb & $\BrMXsb$ \\
\Mllc & $\BrMllc$ &  \MXsc & $\BrMXsc$ \\
\Mlld & $\BrMlld$ &  \MXsd & $\BrMXsd$ \\
\Mlle & $\BrMlle$ &  \MXse & $\BrMXse$ \\
\end{tabular}
\end{ruledtabular}
\end{table}


\begin{thebibliography}{99}
\bibitem{bib:mmiss}
  We define
  $M_{\rm miss}
    =\sqrt{(2\Ebeam-\sum E^{\rm CM}_i)^2/c^4-|\sum{\vec{p}^{\,\rm CM}_i/c}|^2}$ and
  $E_{\rm vis}=\sum E^{\rm CM}_i$,
  where the sum is over all the charged particles with a pion mass hypothesis
  and all photons.
\end{thebibliography}
\end{document}